\documentclass[conference]{IEEEtran}
\IEEEoverridecommandlockouts
\newtheorem{theorem}{Theorem}

\newtheorem{remark}[theorem]{Remark}

\usepackage[noadjust]{cite}

\ifCLASSINFOpdf
  \usepackage[pdftex]{graphicx}
\else
\fi

\usepackage[cmex10]{amsmath}
\usepackage{amssymb}

\usepackage{algorithm}
\usepackage{algpseudocode}

\usepackage{enumitem}

\usepackage[caption=false,font=footnotesize]{subfig}

\usepackage{url}

\begin{document}
\title{Optimality and Rate-Compatibility for Erasure-Coded Packet Transmissions when Fading Channel Diversity Increases with Packet Length}

\author{\IEEEauthorblockN{Sudarsan V. S. Ranganathan, Tong Mu, and Richard D. Wesel}
\IEEEauthorblockA{Department of Electrical Engineering, 
University of California, Los Angeles, Los Angeles, California 90095\\
Email: sudarsanvsr@ucla.edu, tongmu@ucla.edu, wesel@g.ucla.edu}
}


\maketitle
\begin{abstract}
A message composed of packets is transmitted using erasure and channel coding over a fading channel with no feedback. For this scenario, the paper explores the trade-off between the redundancies allocated to the packet-level erasure code and the channel code, along with an objective of a low probability of failure to recover the message. 

To this end, we consider a fading model that we term \textit{proportional-diversity block fading (PD block fading)}. For a fixed overall code rate and transmit power, we formulate an optimization problem to numerically find the optimal channel-coding rate (and thus the optimal erasure-coding rate) that minimizes the probability of failure for various approximations of the problem. 

Furthermore, an interpretation of the results from an incremental redundancy point of view shows how rate-compatibility affects the possible trajectories of the failure probability as a function of the overall code rate. Our numerical results suggest that an optimal, rateless, hybrid coding scheme for a single-user wireless system over the PD block-fading channel should have the rate of the erasure code approach one.
\end{abstract}
\IEEEpeerreviewmaketitle
\section{Introduction}
\label{sec_introduction}
Consider a single-user, point-to-point wireless communication system with the following \textit{hybrid} coding scheme. The transmitter, with a certain number of packets to transmit, codes across the packets using an erasure code; the resultant packets are then transmitted over a channel using a channel code. The receiver decodes each channel packet and then recovers the overall message by decoding the packet-level erasure code. One can implement such a hybrid scheme by using, for instance, ``rateless" erasure codes such as Raptor codes \cite{1638543}, along with powerful channel codes such as low-density parity-check (LDPC) codes \cite{ldpcgall}. 

The hybrid coding scheme can be viewed as a cross-layer coding scheme across the network and physical layers. It can also be considered a physical-layer channel-coding scheme with two layers of coding. In either case, given an overall code rate for the hybrid code, it is not apparent how the transmitter should trade-off the redundancies allocated to the erasure code and the channel code. Moreover, the trade-off depends upon an objective for the system, the channel model, and the hybrid scheme that is implemented. 

This work primarily follows up on the investigations of Courtade and Wesel \cite{5936794} on a generic hybrid coding scheme. The underlying question given an overall code rate is whether such a hybrid scheme necessary or helpful. Hence, we define an objective to be achieved using such a cross-layer interaction. This leads to an optimization problem that will answer the question and show the trade-off between the erasure- and channel-coding rates. Courtade and Wesel \cite{5936794}, for the block-fading channel, obtained results that show the superiority of a hybrid scheme over simple channel coding. They found that, as the overall rate of the hybrid scheme goes to $0$, the optimal erasure-coding rate goes to a non-zero constant less than $1$. Their paper and references therein are all related to our work. 

Other related works include Sun's work \cite{6092888} on a relay-aided system that uses network coding and channel coding. Here, the author solves the redundancy-allocation problem using the block-fading model, where each packet faces only one block fade irrespective of its block-length. Guo et al.\ \cite{6196275} study a similar hybrid scheme in a practical setting using LDPC codes, demonstrating the benefit of a hybrid scheme over others. Koller et al.\ \cite{6089518} also study network-coded unicast and broadcast systems over the binary symmetric channel. They find that longer channel packets are beneficial compared to more redundancy allocated for network coding. 

Hybrid coding schemes \textit{with feedback} have a wholly separate body of literature in different forms of \textit{automatic repeat request (ARQ)}. Heindlmaier and Soljanin \cite{7028505} recently showed that ARQ outperforms hybrid coding for a single-user system, but that for a broadcast system hybrid coding may be superior. 

This paper, in Section \ref{sec_notation}, extends the work of Courtade and Wesel \cite{5936794} to a fading model termed \textit{proportional-diversity block fading (PD block fading)}, where the fading diversity increases proportional to the number of transmitted symbols. By formulating an optimization problem with an objective different from \cite{5936794}, we provide numerical results via Gaussian approximations in Section \ref{sec_solutions_numerical}. Our results show that, for the PD block-fading channel, the optimal erasure-coding rate goes to $1$ as the overall code rate goes to $0$. Section \ref{sec_incremental_redundancy}, which precedes our conclusions, discusses the problem from an incremental redundancy point of view.
\section{Preliminaries and Notation}
\label{sec_notation}
\subsection{Channel Model}
\label{subsec_channel_model}
Consider a transmitter and a receiver, with an antenna each, communicating over a fading channel \cite{goldsmith_wireless}. The channel is modeled as
\begin{align}
\label{eq_fading_channel}
Y = HX + Z,
\end{align}
where $X$ is the transmitted symbol, $Y$ is the received symbol, $H$ is the fading coefficient, and $Z$ is i.i.d.\ additive white Gaussian noise (AWGN) with variance $\sigma^2$ and mean $0$. We assume the following, although similar analysis can be carried out for other channel models:
\begin{enumerate}
\item The channel is Rayleigh fading and $\mathbb{E}\left[H^2\right] = 1$.
\item $Z$ has unit variance, i.e.\ $\sigma^2 = 1$. 
\item The signaling constellation is one-dimensional.
\end{enumerate}
Let the average transmit power be $\mathbb{E}\left[X^2\right]=P$. Then, the instantaneous \textit{signal-to-noise ratio (SNR)} when $H=h$ is $h^2P$. For this Rayleigh fading channel, SNR (denoted $\gamma$) is exponentially distributed with parameter $\frac{1}{P}$ that depends only on the average transmit power. Note that, $\gamma$ has a mean of $P$.
\subsection{Communication System Parameters and Fading Diversity}
\label{subsec_system_params}
A message consisting of $m$ packets with $k$ nats of information per packet is to be transmitted with a low \textit{probability of message error} $q$; this is the probability that the receiver fails to recover all the $m$ packets. The transmitter uses channel \eqref{eq_fading_channel} for $T$ units of time for an overall code rate of $\frac{mk}{T}$. It performs erasure coding across the $m$ packets at a rate $R_E$ and codes each resultant packet at a channel-coding rate $R_C$ such that
\begin{align}
\label{eq_overall_rate}
\frac{mk}{T} = R_E R_C.
\end{align}
That is, the $m$ packets are first coded using an erasure code at rate $R_E$ to yield $\frac{m}{R_E}$ packets. Note that, for erasure coding, $R_E$ has to satisfy $R_E \le 1$. To transmit each packet, the transmitter uses a channel code at rate $R_C$ [nats/channel-use] so that the resultant codeword block-length of each packet is $\frac{k}{R_C}$. For a fixed average transmit power, our objective is to pick the value of $R_C$ (and thus $R_E$) that optimizes an objective function. The unit of channel-coding rate is ``nats/channel-use" for convenience. The receiver is assumed to know the fading coefficient $H$ while the transmitter does not.

 Courtade and Wesel \cite{5936794} assumed the block-fading model in their work. Here, the number of fades, $F$, remains a constant irrespective of the codeword block-length. Refer to Goldsmith \cite{goldsmith_wireless}, Biglieri \cite{biglieri_Coding} for a review of the block-fading model. For our work in this paper, we assume a fading model that we refer to as \textit{proportional-diversity block fading (PD block fading)}. This model is also a block-fading model in that a certain number of channel symbols in a codeword encounter the same fade value $H=h$. But we introduce a parameter $l_f$, a constant that stands for the fade lengths. With the block-length being $\frac{k}{R_C}$, the number of block fades $F_P$ in a transmitted codeword of a system with PD block fading of fade lengths $l_f$ is
\begin{align}
\label{eq_no_fades}
F_P = \left\lceil\frac{k}{R_C l_f}\right\rceil.
\end{align}
With PD block fading, long codewords benefit from an inherent increase in diversity. For this work, we assume that each block-fading event is independent, i.e.\ $H$ assumes i.i.d.\ values across different block fades via \eqref{eq_fading_channel}.

We assume that the receiver decodes the erasure code and recovers the message successfully whenever the channel decoder decodes correctly a subset of the $\frac{m}{R_E}$ packets that it receives (as in \cite{5936794}). The number of packets that the decoder of the erasure code requires to recover the message, denoted $\hat{m} \ge m$, depends upon the erasure code. For Reed-Solomon erasure codes, $\hat{m} = m$; for fountain codes such as a Raptor code, $\hat{m} > m$ typically. Thus, in our analysis we use $\hat{m}$ instead of $m$ as this is a system design parameter that is known, i.e.\ the transmitter has to transmit at least $\hat{m}$ packets implying that $\frac{m}{R_E} \ge \hat{m}$. Also following \cite{5936794}, we assume that the channel codes that are used in the system operate close to capacity with a block-error probability that is assumed to be zero. 

\begin{remark}
\label{remark_1}
Our numerical computations and the ensuing searches of Section \ref{sec_solutions_numerical} can be generalized in a straightforward manner to accommodate any block-error probability for channel codes that may be used in a real system. 
\end{remark}

\begin{remark}
\label{remark_2}
We assume that the codebook for the channel is a Gaussian codebook in our problem formulation in Section \ref{sec_solutions_numerical}. The removal of this assumption for a practical treatment using discrete constellations requires a more careful treatment and this is future work.
\end{remark}
\section{Optimization Problem and Numerical Results}
\label{sec_solutions_numerical}
In this section, we formulate the optimization problem and present approximate solutions along with numerical results from computational searches. For all our results, we restrict the values of $R_C$ so that the number of packets transmitted, $\frac{m}{R_E} = \frac{R_C T}{k}$, is a positive integer.
\subsection{The Optimization Problem}
\label{subsec_optimization}
The receiver of the communication system receives $\frac{m}{R_E} = R_C T k^{-1}$ packets from the channel, of which it has to decode at least $\hat{m}$ packets successfully in order to recover the message. Thus, the probability of message error $q$ can be written using the binomial distribution as
\begin{align}
\label{eq_mep}
q = \sum_{i=0}^{\hat{m}-1} {R_C T k^{-1} \choose i} \left(1-p_e\right)^i p_e^{\left(R_C T k^{-1} - i\right)}.
\end{align}
In the above expression, $p_e$ denotes the probability that a packet is not decoded upon reception from the channel; this is called the \textit{probability of packet erasure}. Owing to our assumption that the channel codes in the system operate close to capacity with zero block-error probability, which also assumes inherently that the block-length $\frac{k}{R_C}$ is long enough, $p_e$ constitutes only one event: \textit{fading outage} \cite{biglieri_Coding}. 

For the block-Rayleigh fading channel with $F \ge 1$ fades that have SNR $\gamma_i\sim \text{Exponential}\left(\frac{1}{P}\right), 1 \le i \le F$, an outage is said to have occurred if the following event takes place \cite{biglieri_Coding}:
\begin{align}
\label{eq_bf}
\left\{\frac{1}{F}\sum_{i=1}^{F}C\left(\gamma_i\right) < R_C\right\},
\end{align}
where $C\left(\gamma_i\right) = \frac{1}{2} \log\left(1+\gamma_i\right)$ is the mutual information of a scalar Gaussian channel that has a Gaussian input and an SNR  $\gamma_i$. This event is the set of all channel realizations along a codeword with $F$ block fades, with an \textit{average mutual information} less than the transmitted code rate $R_C$.

For the PD block-Rayleigh fading model in this paper, the outage event is captured in a similar manner via
\begin{align}
\label{eq_vdbf}
\left\{\frac{1}{\frac{k}{R_C l_f}}\sum_{i=1}^{\left\lfloor\frac{k}{R_C l_f}\right\rfloor}C\left(\gamma_i\right) + \frac{\frac{k}{R_C l_f} - \left\lfloor\frac{k}{R_C l_f}\right\rfloor}{\frac{k}{R_C l_f}}C\left(\gamma_\text{last}\right)< R_C\right\}.
\end{align}
The above event, in its left-hand side, has the \textit{weighted average mutual information} of the $F_P=\left\lceil\frac{k}{R_C l_f}\right\rceil$ fades, and it is a straightforward generalization of \eqref{eq_bf}. 

In order to avoid the corner case of $\left\lfloor\frac{k}{R_C l_f}\right\rfloor = 0$, for brevity, we assume that $R_C$ is bounded as $R_C \le \frac{k}{l_f}$. Note that the ``last" fade with SNR $\gamma_\text{last}$ will take effect only when $\frac{k}{R_C l_f}$ is not an integer. Also, since $R_C \ge \frac{k\hat{m}}{T}$ as $R_E \le \frac{m}{\hat{m}}$, we get another constraint that $\hat{m} \le \frac{T}{l_f}$. Thus, we assume that the input parameters of the optimization problem satisfy
\begin{align}
\label{eq_cons1}
\hat{m} \le \frac{T}{l_f}, ~~~l_f \ll T.
\end{align}

Denote the weighted average mutual information for the PD block-fading channel as $W$, which is the random variable in the left-hand side of \eqref{eq_vdbf}. For the decoder of the channel code, according to the assumption of using capacity-achieving codes, we say that the decoder successfully decodes a channel packet if the event $\left\{W > (1+\epsilon)R_C\right\}$ takes place across the codeword, where $\epsilon$ is a small margin. 

The binomial sum in \eqref{eq_mep} can be computed numerically only for small values of $R_C T k^{-1}$. Hence, we approximate the random variable that denotes the number of packets successfully decoded by the channel decoder using the Central Limit Theorem (CLT), and obtain the Gaussian approximation for $q$ \cite{5936794} as
\begin{align}
\label{eq_q_approx}
q \approx \Phi \left[\frac{(\hat{m}-1) - R_C T k^{-1} (1-p_e)}{\sqrt{R_C T k^{-1} p_e(1-p_e)}}\right],
\end{align}
where $\Phi(x)$ is the value of the cumulative distribution function (CDF) of the standard normal random variable at $x \in \mathbb{R}$. 

\begin{remark}
\label{remark_3}
It must be noted that our work differs from \cite{5936794} in certain key aspects of the system. The fading model in our work provides an inherent diversity to the system that grows with the number of transmitted symbols in a channel packet, whereas \cite{5936794} assumed a block-fading channel, for which the amount of diversity remains constant as the number of symbols in a channel packet grows. Also, in \cite{5936794} the objective is to minimize the transmit power for a tolerable $q$.
\end{remark}

To summarize, the objective of this work is to minimize the message-error probability $q$ in \eqref{eq_mep} via \eqref{eq_q_approx}, where $p_e$ is also a function of $R_C$. Writing the minimization problem in terms of $R_C$, $R_E$ can be obtained as $R_E = \frac{mk}{T R_C}$. Hence, the first instance of the problem with input parameters satisfying \eqref{eq_cons1}, and with the notation and assumptions so far, is the following:
\begin{equation}
\begin{aligned}
\label{eq_mini1}
~~~~\min_{R_C}& ~~ \Phi \left[\frac{(\hat{m}-1) - R_C T k^{-1} (1-p_e)}{\sqrt{R_C T k^{-1} p_e(1-p_e)}}\right], \\
\text{s.t.} ~~ &p_e\left(R_C\right) = \mathbb{P}\left[\frac{1}{\frac{k}{R_C l_f}}\sum_{i=1}^{\left\lfloor\frac{k}{R_C l_f}\right\rfloor}C\left(\gamma_i\right) +\right.\\ &\left.~~~~~~~~~~~~~~~ \frac{\frac{k}{R_C l_f} - \left\lfloor\frac{k}{R_C l_f}\right\rfloor}{\frac{k}{R_C l_f}}C\left(\gamma_\text{last}\right)< (1+\epsilon)R_C\right],\\
&~~~~~~~~~~~ \frac{k\hat{m}}{T} \le R_C \le \frac{k}{l_f}, ~R_C T k^{-1} \in \mathbb{N}.
\end{aligned}
\end{equation}
Note that, minimizing $\Phi(\cdot)$ is equivalent to minimizing its argument, and the value of $q$ need not be explicitly computed. We have specified the dependence of $p_e$ on $R_C$ here for clarity. 

As noted in \cite{5936794}, \cite{5957377}, and many previous works, the evaluation of $p_e$ for the block-Rayleigh fading channel (or for its PD version) is not a straightforward task. One can use \cite{5957377} or similar works for the block-Rayleigh fading channel to compute the outage probability $p_e$ with a minuscule error. But, our fading model complicates it further as we have a sum of two random variables that are not identically distributed in the expression for $p_e$ in \eqref{eq_mini1}. We first expand and rearrange the terms in $p_e$ for our one-dimensional PD block-Rayleigh fading channel with capacity-achieving codes to obtain
\begin{equation}
\begin{aligned}
\label{eq_pe1}
p_e = \mathbb{P}\left[\sum_{i=1}^{\left\lfloor\frac{k}{R_C l_f}\right\rfloor} W_i+ \left(\frac{k}{R_C l_f} - \left\lfloor\frac{k}{R_C l_f}\right\rfloor\right) W_\text{last}< \frac{ck}{l_f}\right],
\end{aligned}
\end{equation}
where $c = 2(1+\epsilon)$, $W_i = \log(1+\gamma_i)$, $W_\text{last} = \log(1+\gamma_\text{last})$. 
\subsection{Gaussian Approximations of the Optimization Problem}
\label{subsec_ga_results}
Now, based on Gaussian approximations of $p_e$ in \eqref{eq_pe1}, as inspired by \cite{5936794}, we present four approximations to the optimization problem \eqref{eq_mini1}. For our numerical-search based results, we pick a very low value of the margin, say $\epsilon=0.05$, to obtain $c$.
\subsubsection{Gaussian Approximation 1 (Approx. 1)}
\label{subsubsec_ga1}
Ignoring the contribution of $W_\text{last}$ in \eqref{eq_pe1}, we get
\begin{align}
\label{eq_ga1_1}
p_e = \mathbb{P}\left[\sum_{i=1}^{\left\lfloor\frac{k}{R_C l_f}\right\rfloor} W_i< \frac{ck}{l_f}\right].
\end{align}
The above can be approximated using the Gaussian CDF as
\begin{align}
\label{eq_ga1_2}
p_e = \Phi \left[  \frac{\frac{c k}{l_f} - \left\lfloor\frac{k}{R_C l_f}\right\rfloor\mu(P)}{\sqrt{\left\lfloor\frac{k}{R_C l_f}\right\rfloor\text{Var}(P)}}  \right].
\end{align}
The values of $\mu(P)$ and $\text{Var}(P)$, which denote the mean and variance of $\log(1+\gamma)$ with $\gamma \sim \text{Exponential}\left(\frac{1}{P}\right)$, can be computed\footnote{$\mu(P) = e^{1/P} \alpha(P)$,\\ $~~~~~\text{Var}(P) = 2e^{1/P} \beta(P) +2e^{1/P} \log(P) \alpha(P) - e^{2/P} \alpha^2(P)$, where \\$~~~~~\alpha(P) = \int_{P^{-1}}^{\infty}\frac{1}{t}e^{-t}dt$ and $\beta(P) =  \int_{P^{-1}}^{\infty}\frac{\log(t)}{t}e^{-t}dt$.} as stated in \cite{5936794}. By ignoring the flooring function, we get \textit{Gaussian approximation 1 (Approx. 1)}, which is an adaptation of (19) in \cite{5936794} to PD block-Rayleigh fading:
\begin{align}
\label{eq_ga1_3}
p_e = \Phi \left[\sqrt{\frac{k}{R_C l_f}}  \frac{c R_C - \mu(P)}{\sqrt{\text{Var}(P)}}  \right].
\end{align}
\subsubsection{Gaussian Approximation 2 (Approx. 2)}
\label{subsubsec_ga2}
For \textit{Approx. 2}, we evaluate \eqref{eq_ga1_2} directly. The approximation to $p_e$ that is being made here is imprecise in the sense that, \eqref{eq_ga1_2} evaluates to the same value for a range of $R_C$ values; the reason being the presence of the flooring function. 

\subsubsection{Gaussian Approximation 3 (Approx. 3)}
\label{subsubsec_ga3}
This approximation is the evaluation of \eqref{eq_pe1} with a constrained search space that limits $R_C$ such that both $\frac{m}{R_E}$ and $\frac{k}{R_C l_f}$ are positive integers. That is, apart from having the number of packets to be transmitted as a positive integer (as we noted before at the beginning of this section), we also assume that diversity can be added only in chunks of one whole fade. This approximation may be required for a system designer, but it severely restricts the search space. Hence, the Gaussian approximation \textit{(Approx. 3)} for this case leads to the same value of $p_e$ as \eqref{eq_ga1_3}; but this equation is more accurate now because of the constraints that are inherent in the optimization problem. 

\subsubsection{Gaussian Approximation 4 (Approx. 4)}
\label{subsubsec_ga4}
The Gaussian approximation that we make here considers both the terms in \eqref{eq_pe1}, making it the most appropriate. Once we find out $\mu(P)$ and $\text{Var}(P)$, we assume that $\sum_{i=1}^{\left\lfloor\frac{k}{R_C l_f}\right\rfloor} W_i$ is Gaussian and also that $\left(\frac{k}{R_C l_f} - \left\lfloor\frac{k}{R_C l_f}\right\rfloor\right) W_\text{last}$ is Gaussian. Thus, their linear sum is another Gaussian random variable denoted $W_G$, which stands for \textit{Gaussian approximation of weighted average mutual information}, with
\begin{equation}
\begin{gathered}
\label{eq_wg_mv}
\text{mean}(W_G) = \frac{k}{R_C l_f} \mu(P), \\
\text{Var}(W_G) = \text{Var}(P) \left[\left\lfloor\frac{k}{R_C l_f}\right\rfloor + \left(\frac{k}{R_C l_f} - \left\lfloor\frac{k}{R_C l_f}\right\rfloor\right)^2\right].
\end{gathered}
\end{equation}
Thus, $p_e$ for this approximation \textit{(Approx. 4)} is
\begin{align}
\label{eq_ga4}
p_e = \Phi \left[  \frac{ \frac{c k}{l_f} - \text{mean}(W_G)}   {\sqrt{\text{Var}(W_G)} }  \right].
\end{align}
\begin{remark}
\label{remark_4}
The Gaussian approximations made above have a few caveats. Obviously, due to CLT, they could be tight enough only when the number of fades for which $p_e$ is being evaluated is large enough. But, we computed the distribution function for $\log(1+\gamma)$ with $\gamma \sim \text{Exponential}\left(\frac{1}{P}\right)$ using \cite{5957377} and observed that, even for one block fade the CDF of outage probability is close to its Gaussian approximation. The results of this comparison are not shown here for brevity.
\end{remark}
\subsection{Computational Results of Numerical Searches}
\label{subsec_results}
The optimization problem \eqref{eq_mini1} is solved here, via a standard brute-force search, for a few values of system parameters via the four Gaussian approximations. The results were obtained using a software implementation in MATLAB. Denote the values of the optimal $R_C$ and $R_E$ as $R_C^*, R_E^*$ respectively. For our results here, we assume that $m=\hat{m}$ for simplicity. 

\begin{figure}[t]
\centerline{\includegraphics[width=3.4in]{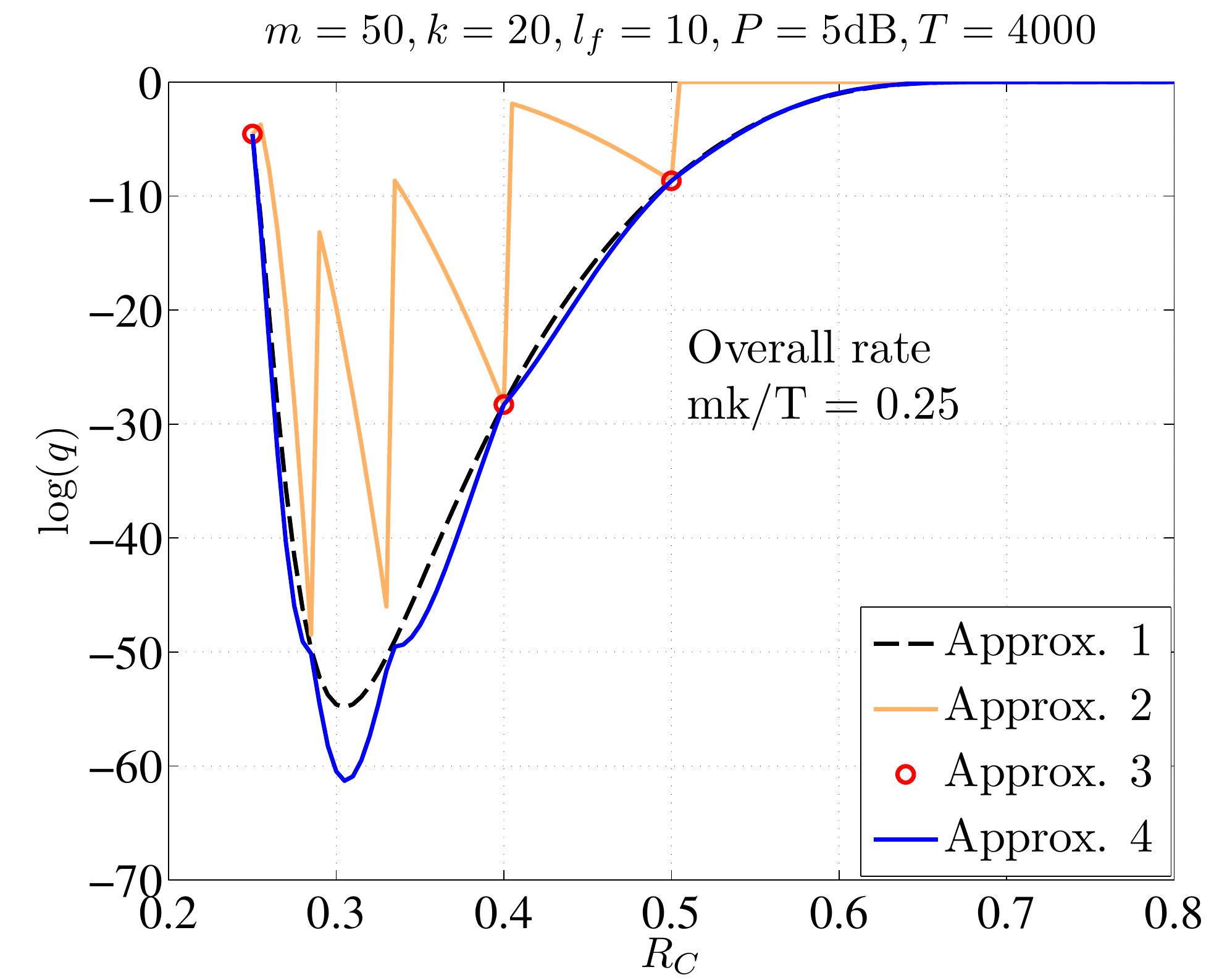}}
\caption{Solution of optimization problem \eqref{eq_mini1} for fixed $mk/T$}
\label{fig_plot1}
\end{figure}

Fig.~\ref{fig_plot1} shows the behavior of $q$ (in $\log$) as a function of $R_C$ in the search space for a fixed overall rate and for a system with $m=50, k=20, l_f = 10$, at a transmit power of $5$dB. The result shows that Approx.\ 1 closely resembles Approx.\ 4, with the latter being our closest approximation of the optimization problem for the PD block-fading channel. Approx.\ 2, which ignores the contribution of $W_\text{last}$, appears ``sawtooth"-like because of the flooring function and the fact that $p_e$ remains constant for a range of $R_C$ values across various ranges in the search space. Approx.\ 3 only has $3$ search points as shown in the plot, where $R_C$ is limited to $0.8$ for clarity. This means that, for a practical system that may add fades in chunks of $l_f$, the optimization problem might be computationally trivial. For future work, we intend to analyze the optimization problem via Approx.\ 1 theoretically as it appears the most tractable; the figure also shows some smoothness for Approx.\ 1. Also, note that Approx.\ 1 and Approx.\ 4 yield almost the same optimal $R_C$ value as shown. 

\begin{figure}[t]
\centerline{\includegraphics[width=3.2in]{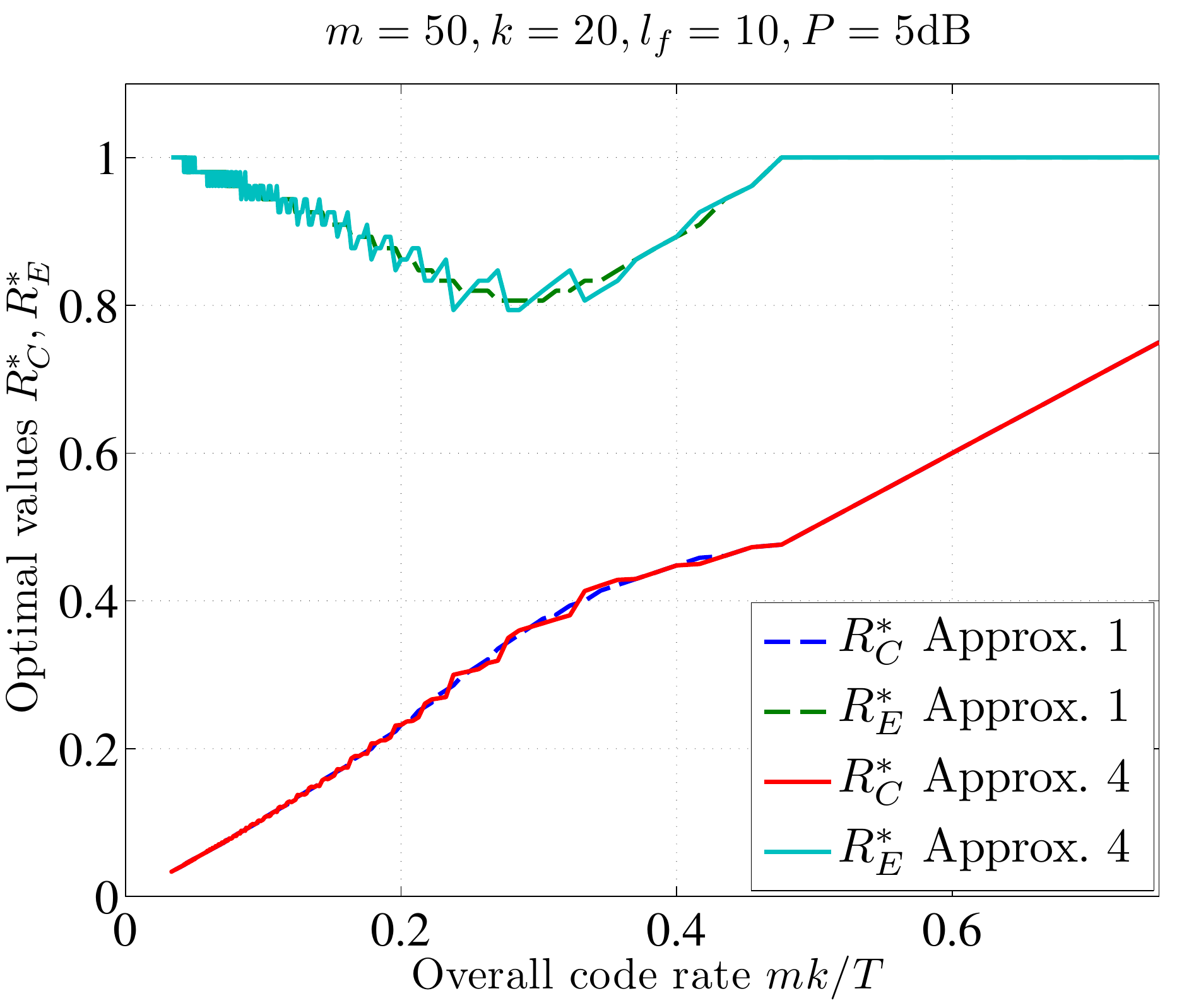}}
\caption{Result of optimization problem \eqref{eq_mini1} as a function of overall code rate}
\label{fig_plot2}
\end{figure}

Fig.~\ref{fig_plot2} shows the optimal values of $R_C$ and $R_E$ obtained from the optimization routine as the overall code rate $\frac{mk}{T}$ goes to 0. We choose to plot only Approx.\ 1 and Approx.\ 4. As observed by Courtade and Wesel \cite{5936794} for the block-fading channel, we also see here for the PD block-fading model that the optimal channel-coding rate goes to $0$. But contrary to their work where the optimal value of $R_E$ approached a non-zero constant less than 1, we see here that it is approaching $1$. This means that, once the overall code rate is low enough, the inherent diversity of the PD block-fading channel causes $q$ to decrease faster compared to the effect of diversity provided by an erasure code. 

\begin{figure}[t]
\centerline{\includegraphics[width=3.42in]{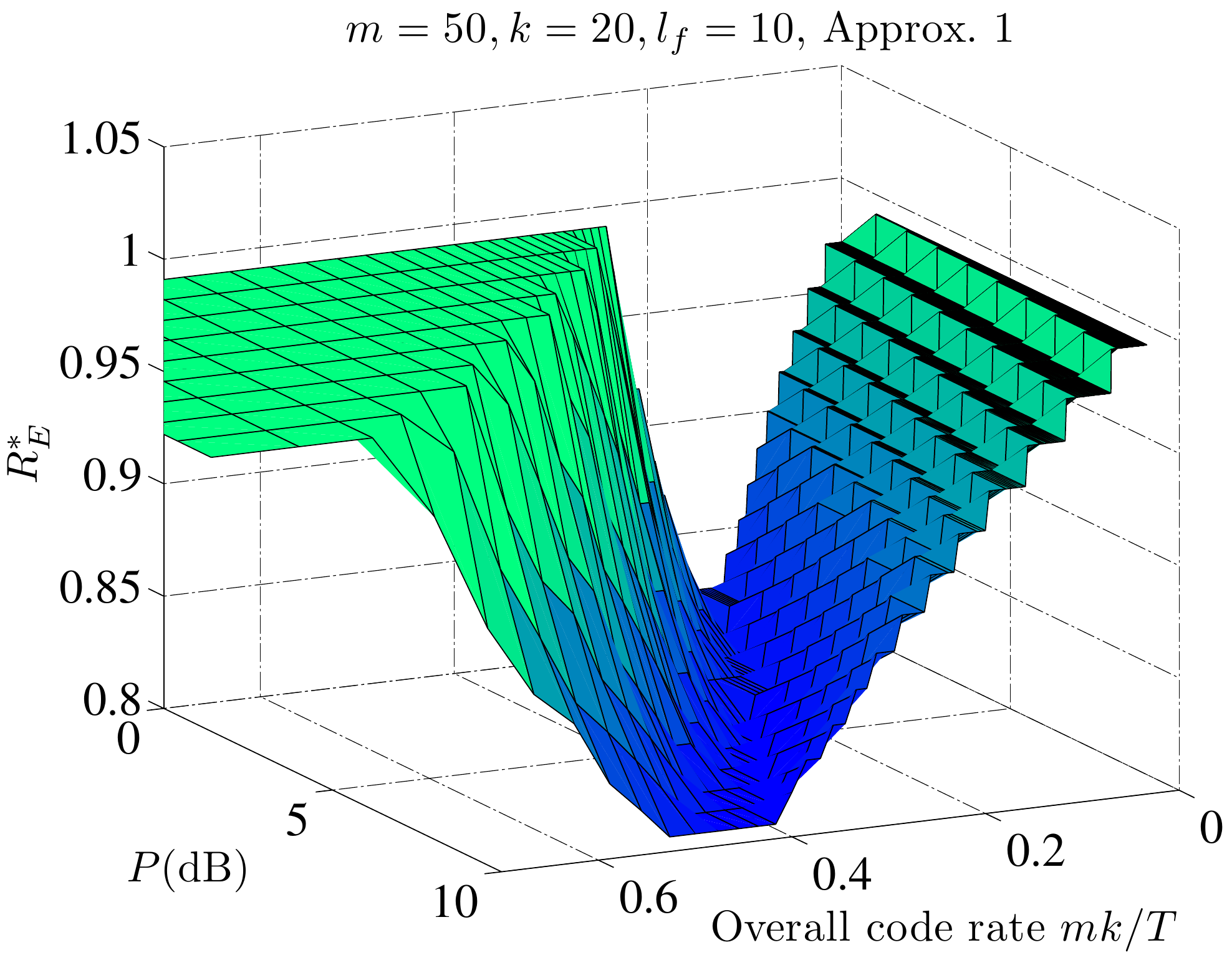}}
\caption{Optimal $R_E$ as a function of overall code rate, for different average transmit power levels $P=1$dB to $P=10$dB in steps of 1dB}
\label{fig_plot3}
\end{figure}

\begin{figure}[h]
\centerline{\includegraphics[width=3.4in]{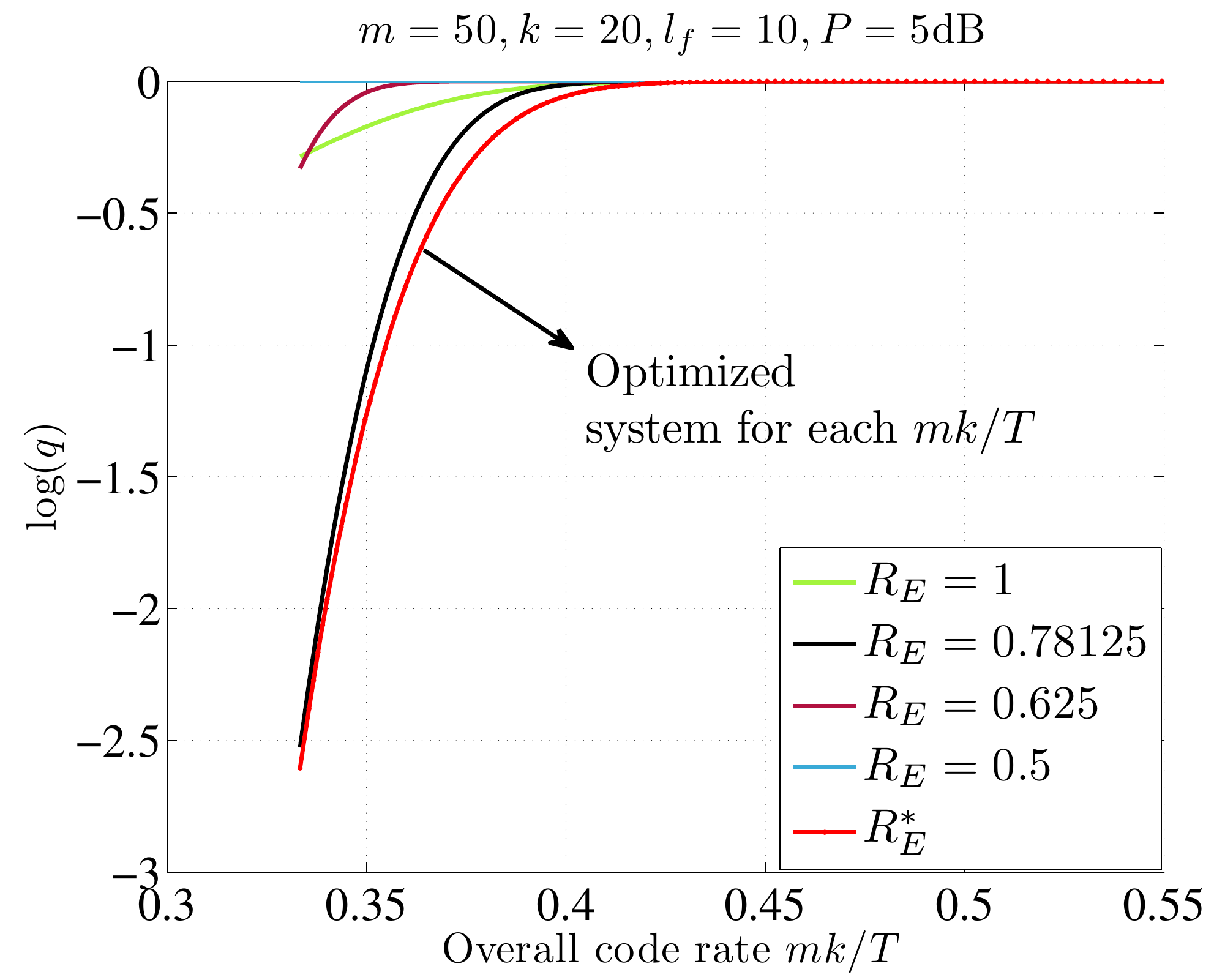}}
\caption{``Ideal" rate-compatible system against rate-compatible systems with fixed $R_E$ via Approx. 4; with hybrid erasure and channel coding}
\label{fig_plot4}
\end{figure}

The same behavior as Fig.~\ref{fig_plot2} was observed for multiple systems, and this is captured in Fig.~\ref{fig_plot3} for the same system as Fig.~\ref{fig_plot2} at different power levels ranging from 1dB to 10dB. The behavior was also observed for systems that varied in the values of $m, k, l_f, T$. 

\begin{remark}
\label{remark_5}
A key aspect of the analysis done so far is that, the values of $k, l_f, T$ can all be scaled by the same amount to indirectly analyze a ``larger'' capacity-achieving system, as the optimization problem \eqref{eq_mini1} remains the same. 
\end{remark}

\section{On Optimal Incremental Redundancy for Hybrid Coding}
\label{sec_incremental_redundancy}
In this brief discussion, we explore one simple opportunity to adapt our hybrid coding scheme to the requirement of incremental redundancy. Such an adaptation is useful in the case of a broadcast setting wherein, as shown  by Heindlmaier and Soljanin \cite{7028505}, hybrid schemes may be superior when compared to ARQ schemes. Assume that the system starts with an initial overall code rate of $m k T^{-1}$. As a best case scenario, assume that the system can potentially add unlimited incremental redundancy once an initial overall code rate is chosen, so that $m k T^{-1} \to 0$. 

From the few numerical results shown so far, observe that the optimal erasure-coding rate of an ``ideal" system with incremental redundancy decreases initially, and then approaches $1$ once the overall code rate is small enough (Figs.~\ref{fig_plot2}, \ref{fig_plot3}). Since this trajectory cannot be followed by a real system, where $R_E$ and $R_C$ can only decrease with time, we leave the question of what the ``optimal'' trajectories of $R_E$ and $R_C$ should be for future work.

In Fig.~\ref{fig_plot4}, we plot the behavior of ($\log$ of) the message-error probability $q$ as a function of the overall code rate. As a simple incremental redundancy scheme, we fix the value of $R_E$ for the system and let $R_C$ decrease with the overall code rate; the chosen values for $R_E$ in the plot are such that $\frac{m}{R_E}$ is an integer. Thus, for every value of the overall code rate shown, there is only one value of $R_C$ and the value of $q$ is obtained via Approx.\ 4. In order to see how suboptimal is this scheme, also plotted is the behavior of $q$ for the ideal system obtained via our computational searches, where for every $T$ we obtain an optimal $R_C, R_E$ pair that minimizes $q$. From the plot, it is clear that fixing a value of $R_E$ can lead to a performance close to an optimal system, and hence one can concentrate on incremental redundancy using only channel coding. But, we see that the system with $R_E = 1$ is far away from the ideal system, and thus a hybrid system with fixed $R_E$ may also suffer a lot if a proper initial value of $R_E$ is not chosen. 
\section{Conclusion}
\label{sec_conclusion}
Following Courtade and Wesel \cite{5936794}, this paper considers the optimal transmission of a hybrid coding scheme with erasure and channel coding. Using a channel model termed proportional-diversity block fading (PD block fading), solving an optimization problem numerically yields the optimal channel- and erasure-coding rates for a fixed overall code rate and transmit power. It is seen contrary to the work of Courtade and Wesel \cite{5936794} that the optimal erasure-coding rate tends to $1$; the primary reason may be attributed to the fact that PD block fading inherently provides increasing diversity to long block-length channel codewords. It is also observed, along the same lines of Courtade and Wesel \cite{5936794} and other works before, that the optimal channel-coding rate goes to $0$ as the overall rate goes to $0$. These results suggest that, with increasing incremental redundancy, one should not make the packet-level erasure codes rateless in a hybrid coding scheme with both erasure and channel coding. 
\newpage
\bibliographystyle{IEEEtran}
\bibliography{IEEEabrv,ICC}
\end{document}